\documentclass[%
 reprint,
superscriptaddress,
5groupedaddress,
 amsmath,amssymb,
 aps,
]{revtex4-2}

\usepackage{graphicx}
\usepackage{dcolumn}
\usepackage{bm}

\usepackage{amsmath,amssymb}
\usepackage[colorlinks=true,bookmarks=false,citecolor=blue,urlcolor=blue]{hyperref} 
\usepackage{braket}
\usepackage{algorithm}
\usepackage{algorithmic}
\usepackage{amsmath}
\usepackage{comment}
\usepackage{soul}
\usepackage{enumitem}

\usepackage{afterpage}

\usepackage{eqnarray}

\begin{document}

\preprint{APS/123-QED}


\title{{Supporting multiple entanglement flows \\
through a continuous-variable quantum repeater}}

\author{Ian J. Tillman}
 \email{ijtillman@email.arizona.edu}
\affiliation{Wyant College of Optical Sciences, University of Arizona, Tucson, AZ 85721, USA}

\author{Allison Rubenok}
 \email{allison.rubenok@arizona.edu}
\affiliation{Wyant College of Optical Sciences, University of Arizona, Tucson, AZ 85721, USA}

\author{Saikat Guha}
 \email{saikat@optics.arizona.edu}
\affiliation{Wyant College of Optical Sciences, University of Arizona, Tucson, AZ 85721, USA}

\author{Kaushik P. Seshadreesan}
\email{kausesh@pitt.edu}
\affiliation{Wyant College of Optical Sciences, University of Arizona, Tucson, AZ 85721, USA}
\affiliation{Department of Informatics and Networked Systems, School of Computing and Information, University of Pittsburgh, Pittsburgh, PA 15260, USA}

\date{\today}

\begin{abstract}

{Quantum repeaters are critical to the development of quantum networks, enabling rates of entanglement distribution beyond those attainable by direct transmission. We consider multiple continuous-variable, squeezed light-based entanglement flows through a repeater involving noiseless linear amplification and dual homodyne detection. 
By analyzing a single-repeater-enhanced channel model with asymmetric half channel losses across the repeater, we determine placements of the central repeater hub in a 4-user hub-and-spoke network that enhance the rate of each entanglement flow.}
\end{abstract}

\maketitle


\section{\label{sec:level1}Introduction}

\noindent The second quantum revolution ~\cite{dowlingMilburnQuantumRevolution} has given rise to novel technologies such as quantum computation~\cite{quantumComputationReview}, quantum sensing ~\cite{quantumSensingReview} and quantum cryptography~\cite{Pirandola:20,scaraniQKDReview,peng2022satellite}. 
There is a growing interest in forming networks of quantum computers and sensors for distributed applications. 
Developing quantum networks and interconnecting them to form a global-scale ``quantum internet” ~\cite{wehnerQuantumInternet,kimbleQuantumInternet} requires reliable quantum communication over long-haul interconnecting links. 
Due to their ability to be transmitted over vast distances photons are a natural choice of information carrier.
Consequently, the primary challenge to overcome in realizing quantum networks and the quantum internet is photon loss in the form of fiber losses, free space attenuation, coupling losses, and detector inefficiencies.

In classical communications, data rate deterioration due to photon loss can be mitigated using electrical regenerators or optical amplifiers at intermediary nodes. 
In quantum communications, the directly analogous deterministic quantum-limited amplifiers (both phase insensitive and phase sensitive) are ineffective in mitigating the effects of photon loss~\cite{namikiNoRegeneration}. 
Thus, special purpose quantum regenerative repeaters, in the form of quantum processors equipped with optical sources, detectors and quantum memories have been proposed to `effectively' amplify quantum information-bearing signals and extend the range of communication. 
A variety of different quantum repeater architectures have been analyzed. They differ in the type of quantum information encoding~\cite{DVReview,gaussianQuantumInformation}, or in the type of the underlying protocols~\cite{repeaterGenerations}.



For the so-called continuous variable (CV) quadrature-based bosonic encodings~\cite{gaussianQuantumInformation} such as the coherent and squeezed states of light, there are several proposed quantum repeater architectures~\cite{wu2022continuous,kaushikPRR}.
One proposed architecture involves the use of the quantum scissor (QS)~\cite{peggQuantumScissor}.
The QS is a non-deterministic operation that, when successful, 
approximates the action of a noiseless linear amplifier (NLA) on low mean photon number states ~\cite{ralphLundQSNLA}. 
In CV entanglement distribution based on transmission of two-mode squeezed vacuum (TMSV) light from spontaenous parametric downconversion, the QS acting on the lossy transmitted mode can probabilistically herald high purity states of higher entanglement, in terms of the logarithmic negativity and entanglement of formation than the original lossy TMSV state \cite{diasRalph2018}. 
In particular, it is possible to tune the gain parameter of the QS such that the heralded states have distillable entanglement~\cite{horodeckiEntanglementMeasures} exceeding the direct transmission entanglement distribution capacity of the channel, $C_{\mathrm{direct}}$, which can be expressed as~\cite{PLOB}
$$ C_\mathrm{direct} = -\mathrm{log}_2(1 - \eta)\ \mathrm{ebits/mode} $$
at any given transmission distance, where $\eta$ is the transmissivity of the channel~\cite{kaushikPRA}. 
Such distilled entangled states shared over two adjacent quantum links can then be used to extend the range of entanglement via entanglement swapping. 
The QS has been shown to support quantum repeater action for entanglement distribution when used with TMSV light sources, mode multiplexing, and multi-mode quantum memories~\cite{diasCVRepeater,kaushikPRR, furrerAndMunro}. 
The entanglement distribution rate vs end-to-end distance has been determined also for the limiting case of an ideal NLA consisting of infinitely many QSs~\cite{ghalaiiInfiniteQS}. 

\begin{figure*}
    \centering
    \includegraphics[width=\linewidth]{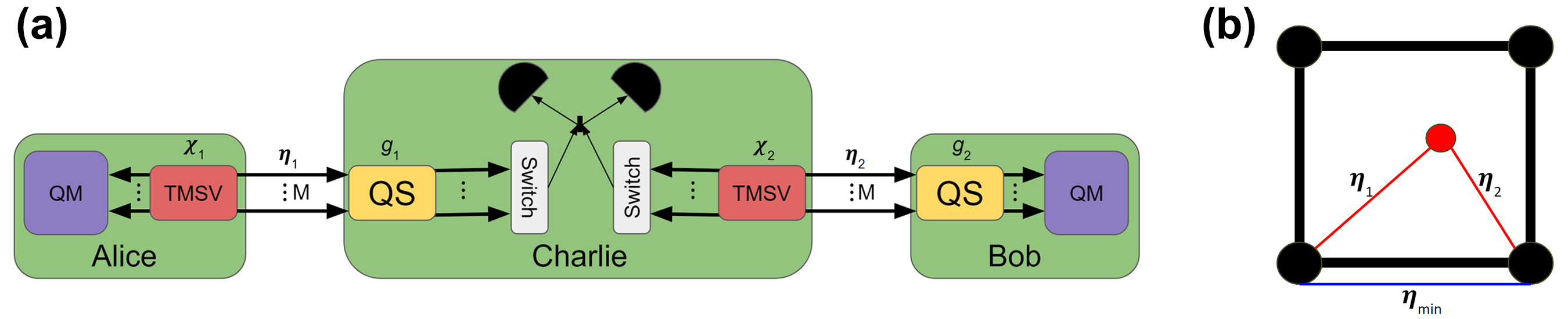}
    \caption{\textbf{(1a)} We model two-user channels based on two sets of asymmetric half channels, each comprises a two-mode squeezed vacuum (TMSV) source and a quantum scissor (QS) separated by a pure loss channel. One mode from each half channel is mixed using dual homodyne detection (DHD) resulting in the two-user channel. \textbf{(1b)} Our network model consists of four users, each possessing the necessary equipment to function as either 'Alice' or 'Bob', placed on the vertices of a square. They all share a single repeater node ‘Charlie’ placed somewhere in the center of the square. We compare the rates we achieve to the repeaterless capacity for both the shortest path between the users (blue) and the path going through our Charlie node (red).
}
    \label{fig:setup}
\end{figure*}

We consider the CV repeater architecture based on multiplexed TMSV state transmissions, QS-based entanglement distillation and coherent dual homodyne detectiton (DHD)-based entanglement swapping in the context of a hub-and-spoke network. 
The network consists of quantum links between four end users and the central repeater hub node, with DHD being performed at the hub between any two links.
Our work considers for the first time a generalized repeater-enhanced channel model with non-symmetric parameters (i.e., transmissivity, scissor gain, and squeezing amplitude) for the two links being connected by the repeater node.
This more general model allows us to explore arbitrary placements of our hub repeater node that could cater to multiple entanglement flows between different pairs of end users, creating a non-trivial optimization problem.
Moreover, the model also corresponds to and enables us to analyze more realistic real-world network scenarios than previously considered grid networks with regular spacings. 
Using provable lower bounds on the distillable entanglement of the final two-mode state, we find placements for which the repeater hub can help surpass the repeaterless capacity for
each entanglement flow within a square network. 
Our work paves the way towards implementing repeater-enhanced distributed sensors~\cite{Xia_distri_sensing, Xia_experiment,Optimal_distri_sensing,Zhang_2021} and long baseline telescopes~\cite{Lukin_19,Gottesman_12}, which are important applications of the quantum internet.

The paper is organized as follows. In section ~\ref{model} we explain our repeater model and motivate a definition for entanglement rate. Section ~\ref{results} contains our results. In Section ~\ref{discussion} we discuss nuances in our model, summarize, and look at future directions for this work.

\section{Repeater Model and Rate Formula}
\label{model}

Our model of a two-user repeater-enhanced end-to-end quantum channel {with non-symmetric parameters} for entanglement distribution is shown in Fig.~\ref{fig:setup} (a). 
It consists of two sets of multiplexed, adjacent ``half channels", one from each of the two users to a repeater node between them.
Each half channel consists of a TMSV source, pure-loss transmission {of arbitrary transmissivity} of one of two modes in the TMSV state, and a QS-based approximate NLA. 
A successfully heralded half channel from each of the two sets is picked with the help of a switch and the two are connected by a coherent DHD entanglement swap operation at the repeater node, resulting in the two-user end-to-end channel.
Below we mathematically describe each of these elements.

The TMSV state is traditionally described in terms of the mode annihilation operators, $\hat{a}$ and $\hat{b}$, and the squeezing operator $S(\zeta) = e^{\zeta^* \hat{a}\hat{b} - \zeta \hat{a}^\dagger \hat{b}^\dagger}$ acting on the joint vacuum state $\ket{0,0}_{AB}$. 
We find it useful to expand this in the Fock basis:
\begin{align}
    \ket{\chi}_{AB}\equiv S(\zeta)\ket{0,0}_{AB} = \sqrt{1-\chi^2} \sum_{n=0}^{\infty} \chi^n \ket{n,n}_{AB},
\end{align}
we take $\zeta \geq 0$ and define $\chi \equiv \mathrm{tanh}(\zeta)$ so that $0 \leq \chi < 1$. 
When one of the two modes from the TMSV state is transmitted through a pure loss bosonic channel of transmissivity $\eta$, the distillable entanglement of the distributed state in the limit of infinite squeezing attains the capacity of the underlying lossy bosonic channel for entanglement distribution. 
This can be shown by calculating the reverse coherent information (RCI)-based Hashing lower bound on the distillable entanglement of the lossy TMSV state, which in limit of $\chi \rightarrow$ 1 attains its maximum value that matches the capacity $C_{\textrm{direct}}$~\cite{PirandolaRCI}.

The action of an ideal NLA on a generic quantum state $|\psi\rangle$ is defined as
\begin{align}
    \ket{\psi} \equiv \sum_{n=0}^\infty c_n \ket{n} \mapsto \hat{T}_\infty \ket{\psi} = A\sum_{n=0}^\infty g^n c_n \ket{n}
\end{align}
where $\hat{T}_k$ sends Fock state $\ket{n} \mapsto g^n \ket{n}$ if $n \leq k$ and removes all higher order modes. 
The QS, as shown in Fig. 1b of ~\cite{diasCVRepeater} and experimentally realized in~\cite{ralphExperiment}, is an operation based on single photon injection and detection that can probabilistically herald an approximation of this transformation for low mean photon number states. 
The QS operation truncates the input quantum state to only its $\{0,1\}$ Fock state support, while amplifying its $1$ photon component relative to the vacuum component. 
This is mathematically described as:
\begin{align}
    \ket{\psi} \mapsto \hat{T}_1 \ket{\psi} = A\left( c_0 \ket{0} + g c_1 \ket{1}\right),
\end{align}
and succeeds with probability
\begin{align}
    P_{\mathrm{NLA}} = \frac{(1-\chi^2)(\chi^2 (\eta g^2 + \eta - 1) + 1)}{(1 + g^2) ((\eta - 1)\chi^2 + 1)^2}
\end{align}
where A is a normalizing constant and $g$ is called the NLA gain. 

Coherent DHD involves mixing two modes on a 50:50 beam splitter followed by measurement of two conjugate quadratures on the two modes. 
When performed on one mode from each of two half channels at a repeater it accomplishes entanglement swapping, resulting in conditional long range entanglement between the other unmeasured modes of the two half channels.
The DHD is depicted in the center of Fig.~\ref{fig:setup} (a). 
The mixing and measuring can be modeled by projecting these two modes onto the eigenstates defined by the complex measurement outcome $\gamma$~\cite{diasCVRepeater}:
\begin{align}
    \ket{\gamma}_{FC} = \frac{1}{\sqrt{\pi}}\sum_{n=0}^\infty \hat{D}_C(\gamma) \ket{n}_C \ket{n}_F = \\ \nonumber \frac{1}{\sqrt{\pi}}\sum_{n=0}^\infty e^{-|\gamma|^2/2}e^{\gamma \hat{c}^{\dagger}} e^{-\gamma^* \hat{c}} \ket{n}_C \ket{n}_F
\end{align}
where the modes $F$ and $C$ are the modes being mixed and measured with DHD. The mode notation is described in Appendix \ref{appendix:a}.

\begin{figure}
    \centering
    \includegraphics[width=0.99\linewidth]{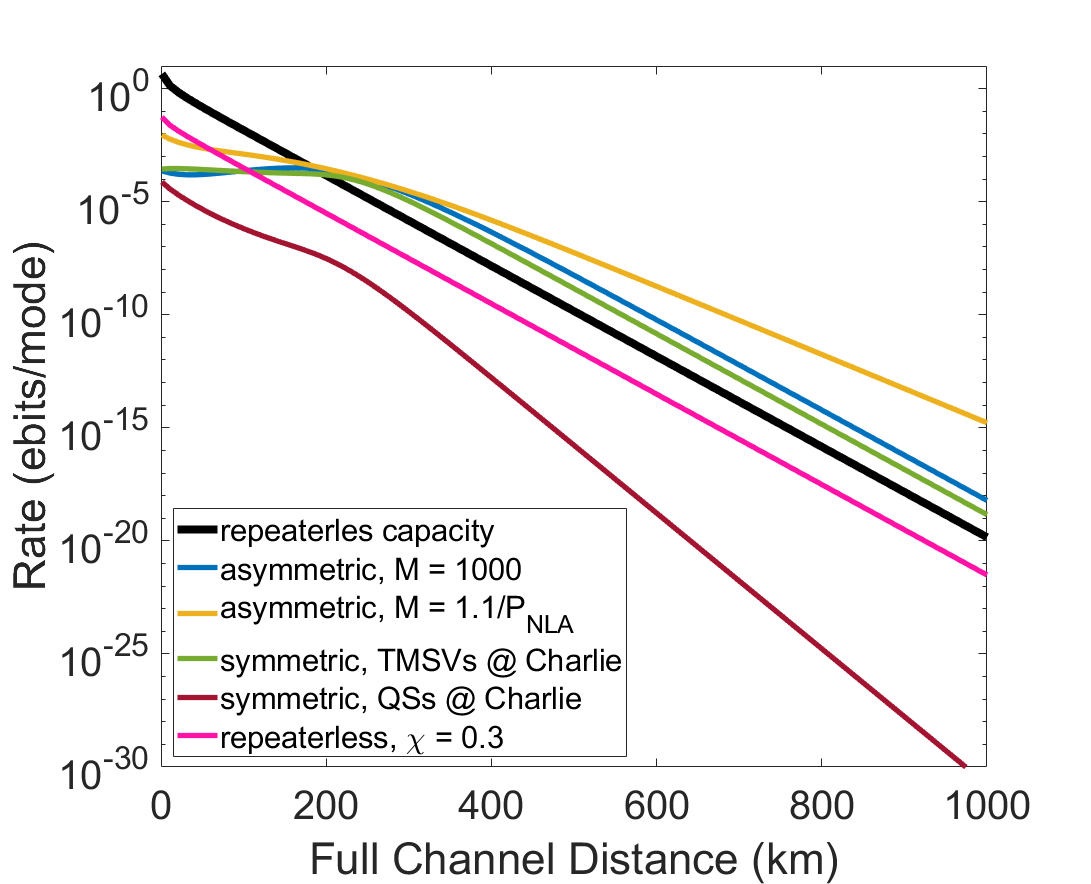}
    \caption{Comparison of all setups. The asymmetric distribution case is optimal, attaining better-than-repeaterless-capacity scaling when we allow optimizing the multiplexing factor, $M$, at every distance.}
    \label{fig:ebitratevsdistance}
\end{figure}

In multiplexed quantum repeaters ~\cite{saikatRateLoss,kaushikPRR}, where one successful half channel is heralded from each set of multiplexed half channels over adjacent quantum links, a lower bound on the per-mode end-to-end entanglement distribution rate can be calculated in terms of the probability that at least one half channel per link succeeds, the probability of successful entanglement swap, the multiplexing $M$, and the RCI of the conditional end-to-end entangled state. 
For the two-user channel considered in this work, where they are connected through a single repeater node, given identical half channel success probabilities $p$ over the two sets of half channels, entanglement swap success probability $q$, multiplexing $M$ and end-to-end entangled state $\rho$, the rate lower bound is given by
\begin{align}
    R = \frac{I_{AB}(\rho) \times q\times (1 - (1 - p)^M)^2}{M} \textrm{ebits/mode}
    \label{eq: Rate_eqn}
\end{align} 
where $I_{AB} (\rho)$ is the RCI. Because DHD is a deterministic measurement, we use $q=1$ in this work.
{The improvement due to multiplexing comes from the ratio $(1-(1-p)^M)^2/(Mp^2)$ going above 1 when $1 <  M \lessapprox 1/p^2$ and $p \ll 1$.
This ratio peaks at about $0.4/p$ when $M \approx 1/p$. 
For the asymmetric half channel based on TMSV and QS with a gain $g$ chosen from the power law discussed in \cite{kaushikPRR}, in the regime $ \eta \ll 1 $ we have $p=P_{\mathrm{NLA}} \sim \eta^{1/4}$. Because $I_{AB}(\rho) \sim \eta$, this implies $ R_{\mathrm{opt}} \sim I_{AB}(\rho) \eta^{-1/4} \sim \eta^{3/4}$, as shown in Fig. \ref{fig:ebitratevsdistance} as the yellow curve. 
However with greater distance this requires exponentially more multiplexed channels to achieve the optimal entanglement distribution rate per mode.}

For a two-user channel between Alice (A) and Bob (B) obtained by connecting two half channels of, {in general, non-symmetric parameters} at a repeater node (C) by DHD (with outcome $\gamma$), we determine the exact form of the density operator of the conditional end-to-end quantum state $\rho_{AB}(\gamma)$. 
We numerically approximate this by truncating it in the Fock basis to a point where the trace is 99\% of the theoretical trace. The density operator $\rho_{AB}(\gamma)$ is normalized as $\frac{\rho_{AB}(\gamma)}{\mathrm{Tr}\left[ \rho_{AB}(\gamma) \right]}$, where the normalization (denominator) is the unnormalized probability distribution of $\gamma$. 
The eigenvalues of both $\rho_{AB}(\gamma)$ and $\rho_A(\gamma)$ are numerically determined for every $\gamma$ in the complex plane that gives a non-negligible RCI and used to evaluate the RCI lower bound.
Using $\mathrm{Tr} \left[ \rho_{AB}(\gamma) \right]$ for each $\gamma$, the entanglement rate is averaged over the $\gamma$ distribution to give us an achievable entanglement distribution rate which we call the 'ergodic' rate~\cite{kaushikPRA}. Details are in Appendix \ref{appendix:a}.

\begin{figure*}
    \centering
    \includegraphics[width=\textwidth]{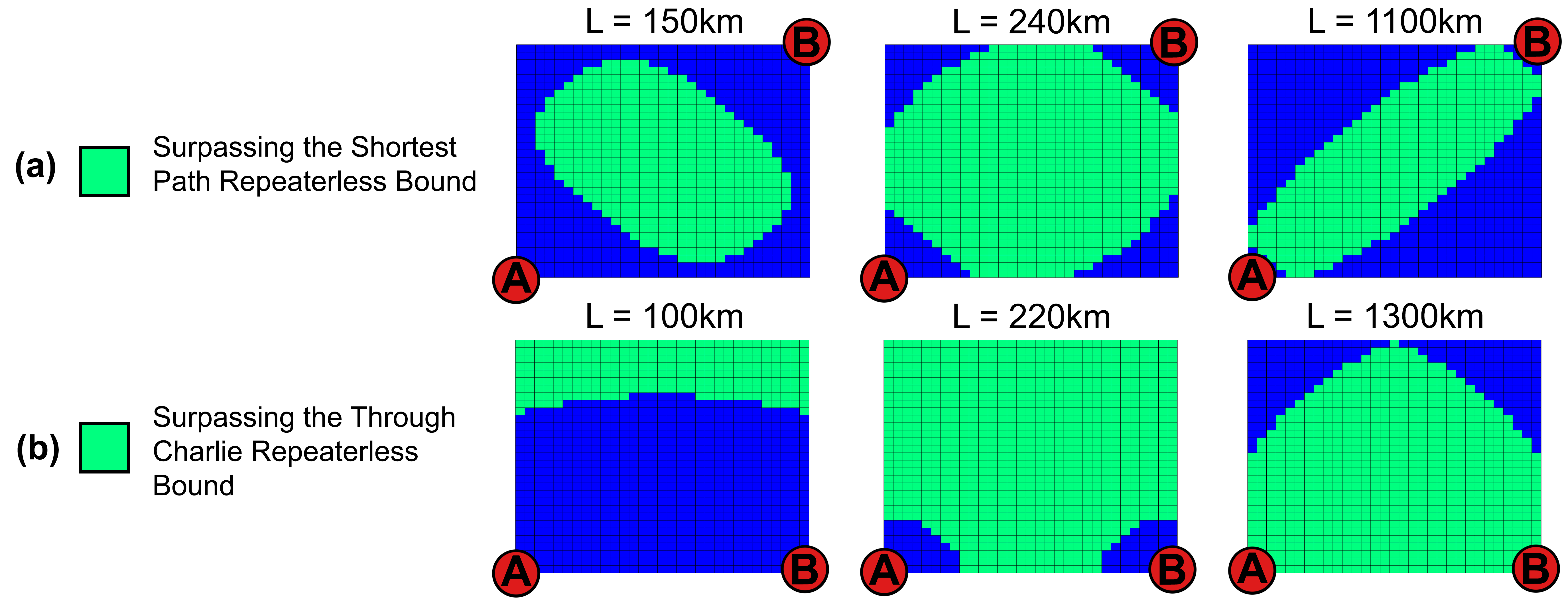}
    \caption{\textbf{(3a)} The area within the square beating the shortest path repeaterless capacity between two diagonal users. The size of the area in which we beat the repeaterless capacity grows to some maximum and then shrinks. \textbf{(3b)} The area beating the through Charlie repeaterless capacity for two side users.}
    \label{fig:example}
\end{figure*}

\section{Results}
\label{results}
{We analyze the end-to-end entanglement distribution rate (per mode, calculated according to~(\ref{eq: Rate_eqn})) vs distance for a two-user channel for all three possible different orientations of the quantum links: i) the setup shown in Fig. \ref{fig:setup}(a) where the users have different equipment, ii) a setup where both users have a QS, and iii) a setup where both users have a TMSV source. } 
For the case of half channels with symmetric transmission losses, Fig. \ref{fig:ebitratevsdistance} plots the rate-distance tradeoff for the different orientations of half channels; two of these are found to beat the repeaterless capacity with $M=1000$. {This figure also shows that optimizing $M$ at each distance without an upper bound for the asymmetric orientation leads to better scaling than the repeaterless capacity.} Though this optimization is trivial, we do not include it in our network entanglement rate analysis because doing so is infeasible for a physical implementation.
Among the different orientations, the one where one end user has a TMSV source (Alice) and the other has a QS-based approximate NLA (Bob), while the intermediate repeater node (Charlie) has one of each, as shown in Fig.~\ref{fig:setup} (a) yields the highest rates. All results shown subsequently are for this scenario with $M=1000$.

\begin{figure*}
    \centering

    
    \includegraphics[width=0.8\textwidth]{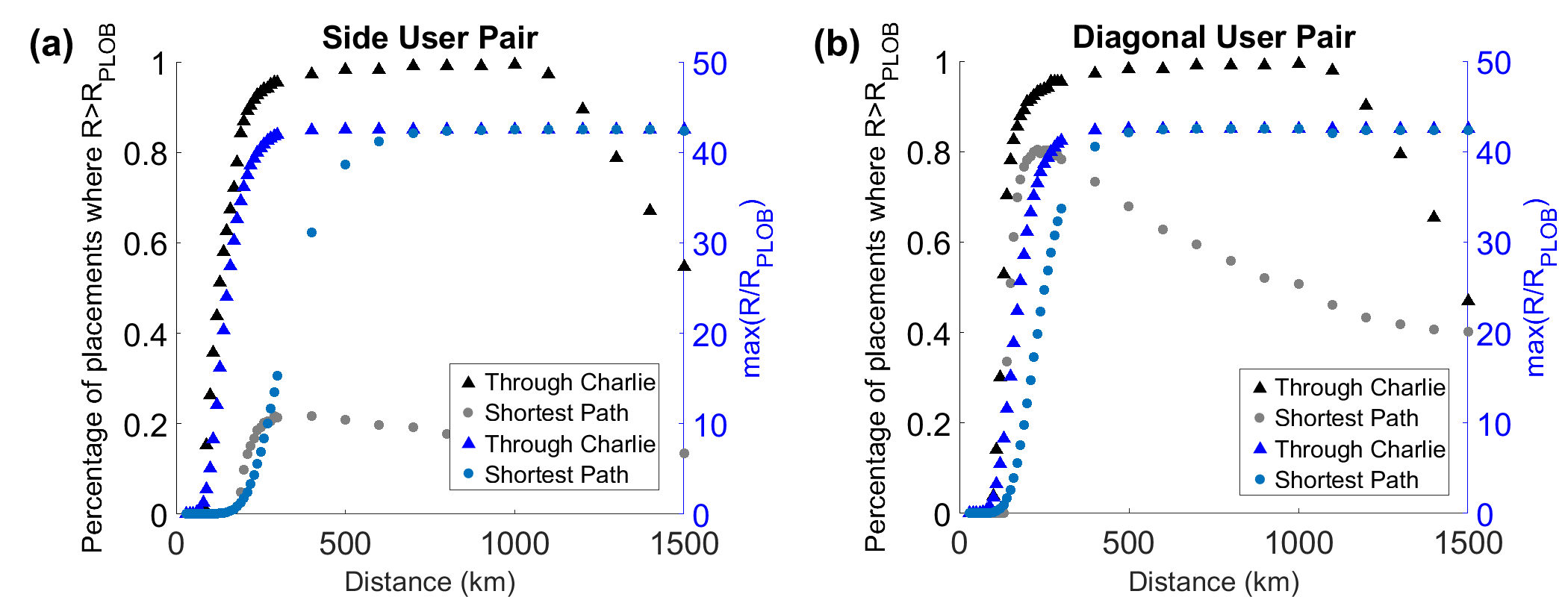}

    \caption{\textbf{(4a - 4b)} The proportion of area beating repeaterless capacity for both a side and diagonal connection. We see an initial increase followed by a decrease once the asymmetric losses lower our rate too much.} 
    \label{fig:plots}
\end{figure*}

For any placement of Charlie (defined by $\eta_1,\eta_2$) we can optimize the TMSV and NLA parameters ($\chi_1,\chi_2,g_1,g_2$) to maximize the entanglement rate. For brevity we choose not to optimize over $\chi_1,\chi_2$, which we set to 0.3 universally, and we quasi-optimize $g_1,g_2$ by the power law discussed in \cite{kaushikPRR} using the respective half channel losses. This gives us noticeable improvements over fixing $g_1,g_2$.

Now that we can approximate the optimal achievable rate for a general {single repeater-enhanced end-to-end} channel, we set up our 4-user network and analyze the rates. 
Our network model is depicted in Fig.~\ref{fig:setup} (b), consisting of four users with the one central hub repeater node, `Charlie', that can act as a repeater for any two users wanting to communicate.
We compare the rates to three baselines: 
\begin{enumerate}[label=(\alph*)]
    \item Repeaterless capacity through the intermediary node
    \item Repeaterless capacity through the shortest possible path between the nodes
    \item Rate achieved using only direct DHD at an end user.
\end{enumerate}

\noindent We mainly focus on baseline (a), since for long distance scenarios we envision a practical network will be planned such that all users have access to at least one shared repeater node facilitating communication between them, where as only a smaller subset of nodes will have the possibility to communicate via a direct connection. (c) can be seen as an experimentally possible outcome as opposed to the purely theoretical upper limits of (a) and (b). We set the squeezing parameter for (c) to be $\chi = 0.3$ to match our modeled TMSV sources used in the repeater links, but we use the optimal setup of assuming the DHD swap occurs at an end user instead of at the repeater node.

We looked at the percentage of placements within the square network where our repeater node surpasses the baselines listed above for diagonal and adjacent user pairs. 
For every network scale we split the network into a 31x31 grid and calculated the rates of each user pair for every Charlie placement on that grid assuming standard 0.2 dB/km loss. 
Examples comparing our rates to (a) and (b) are shown in Fig. \ref{fig:example}. How often we beat these benchmarks in the network scenario is shown in Fig. \ref{fig:plots}. 
Fig \ref{fig:example} gives some intuition as to why the curves in Fig. \ref{fig:plots} initially rise and then drop back down again after recognizing that our rates are lowered in cases of asymmetric loss. 
We note that in the case of baseline (a), placing Charlie at the center of the network will always service any user pair better than repeaterless capacity for network scales longer than about 200 km, however in the case of baseline (b) there is no placement at any scale that surpasses repeaterless capacity for all user pairs. 
With our numerics we were able to service all user pairs better than benchmark (b) only after placing a minimum of 3 Charlie nodes within the network. 
{With respect to both baselines (a) and (b),} we beat the repeaterless capacity by up to a factor of about 40 (as seen in Fig. \ref{fig:plots}) and the direct DHD performance {(baseline (c))} by up to a factor of about 1750. These correspond to the vertical distances between the asymptotically-parallel lines in Fig. \ref{fig:ebitratevsdistance}.

\section{Discussion}
\label{discussion}

We chose to calculate the so-called `ergodic' rate rather than the rate achieved by the average state because it is higher than the latter while remaining achievable. The ergodic rate can in principle be attained by considering independent asymptotic entanglement distillation protocols implemented over multiple infinitesimally small bins spanning the complex plane. 

To assess when such CV repeater enhanced networks will be experimentally viable future work should seek to: fully optimize the half channel squeezing and gain parameters, cap the gain parameters at some $g_{\mathrm{max}}$, and consider a more realistic model for the half channel inclusive of added thermal noise from the environment. Currently, the highest recorded gain approximating an NLA is 12 \cite{ralphExperiment}. In the absence of thermal noise, we begin to see an improvement in the repeaterless capacity at a gain of 13.

Another consideration is the minimum equipment required to access the network. We have found rates in our network to be optimal when one end user has a TMSV and the other has a QS. Therefore in order for all pairs to be able to communicate, every user will need to be in possession of both a TMSV source and a QS, in contrast to the symmetric case where users only need one of the two. 

By doing an RCI analysis of a generalized link model with non-symmetric parameters, we have shown that a QS acting as an approximate NLA can be used in quantum repeaters that surpass the repeaterless capacity at large distances. Within our 4-user network we surpass repeaterless capacity through Charlie for all user pairs with a single Charlie placed in the center of the square for side lengths greater than 200 km. 
Using a model with non-symmetric half channel losses allowed us to show that this setup is robust enough to asymmetries in loss to be used as a shared repeater node in a network while allowing most, if not all, user pairs to surpass the repeaterless capacity simultaneously. Our work paves the way to realizing realistic CV repeater networks, including distributed sensing and computing applications as well as quantum key distribution over the future quantum internet.

This work was supported by NSF grants 1842559 and 2204985.

\bibliography{PRA}
\bibliographystyle{unsrt}

\appendix

\section{Density Matrix Calculations}
\label{appendix:a}

We begin with the states in \cite{diasCVRepeater} and follow much of the same algebra. We label the modes implied in Fig. \ref{fig:setup} as follows:

\begin{itemize}
    \item A $\rightarrow$ Alice's stored mode
    \item B $\rightarrow$ Bob's stored mode
    \item C $\rightarrow$ the mode between Charlie's QS and the DHD
    \item D $\rightarrow$ the environment mode on the link between Alice and Charlie
    \item E $\rightarrow$ the environment mode on the link between Bob and Charlie
    \item F $\rightarrow$ the mode between Charlie's TMSV source and the DHD
\end{itemize}

This gives us the following independent half channel states just before the DHD, written in the Fock basis:

\begin{align}
    \ket{\psi}_{ACD} &= \sqrt{\frac{1-\chi_1^2}{1+g_1^2}} \Big[ \sum_{n=0}^\infty \chi_1^n \alpha_1^n \ket{n,0,n}_{ACD} \nonumber\\
    &+ g_1 \sqrt{\eta_1} \sum_{n=1}^\infty \chi_1^n \sqrt{n} \alpha_1^{n-1} \ket{n,1,n-1}_{ACD} \Big ]
\end{align}

\begin{align}
    \ket{\psi}_{BEF} &= \sqrt{\frac{1-\chi_2^2}{1+g_2^2}} \Big[ \sum_{n=0}^\infty \chi_2^n \alpha_2^n \ket{0,n,n}_{BEF} \nonumber\\
    &+ g_2 \sqrt{\eta_2} \sum_{n=1}^\infty \chi_2^n \sqrt{n} \alpha_2^{n-1} \ket{1,n,n-1}_{BEF} \Big],
\end{align}

\noindent where $\eta_1,\eta_2$ are the half channel transmissivities, $\chi_1,\chi_2$ are the squeezing parameters for Alice and Bob, $g_1,g_2$ are the QS gains on Alice and Bob's half channels, and $ \alpha_k \equiv \sqrt{1 - \eta_k}$. 
Taking the inner product of the tensor product of these two with the measurement eigenstate
\begin{align}
    \nonumber \ket{\gamma}_{FC} = \frac{1}{\sqrt{\pi}}\sum_{n=0}^\infty \hat{D}_C(\gamma) \ket{n}_C \ket{n}_F = \\ \frac{1}{\sqrt{\pi}}\sum_{n=0}^\infty e^{-|\gamma|^2/2}e^{\gamma \hat{c}^{\dagger}} e^{-\gamma^* \hat{c}} \ket{n}_C \ket{n}_F
\end{align}
\noindent corresponding to dual homodyne detection on modes $F$ and $C$, and tracing out the $D$ and $E$ environment modes, we obtain the final conditional density operator $\rho_{AB}(\gamma)$ of modes $A$ and $B$ (expanded in the Fock basis) given by

\begin{widetext}

\begin{eqnarray}
\begin{aligned}
    \rho_{AB} &= \frac{e^{-\gamma^2}}{\pi} \left( \frac{1-\chi_1^2}{1+g_1^2} \right) \left( \frac{1-\chi_2^2}{1+g_2^2} \right)
    \\
    &\quad \Bigg[ \sum_{n=0}^\infty \sum_{m=0}^\infty \chi_1^{2n} \alpha_1^{2n} \chi_2^{2m} \alpha_2^{2m} \frac{(-\gamma)^{2m+1}}{m!} \ket{n,0}\bra{n,0}_{AB}
    \\
    &\quad + \sum_{n=0}^\infty \sum_{m=0}^\infty g_2 \sqrt{\eta_2} \chi_1^{2n} \alpha_1^{2n} \chi_2^{2m+1} \alpha_2^{2m} \frac{(-\gamma)^{2m}}{m!} \ket{n,0}\bra{n,1}_{AB}
    \\
    &\quad + \sum_{n=0}^\infty \sum_{m=0}^\infty \sum_{k=\mathrm{max}(m-1,0)}^m g_1 \sqrt{\eta_1} \chi_1^{2n+1} \alpha_1^{2n} \chi_2^{2m} \alpha_2^{2m} \sqrt{n+1} \frac{(-\gamma)^{m+k}}{k!}\gamma^{1+k-m} \ket{n,0}\bra{n+1,0}_{AB}
    \\
    &\quad + \sum_{n=0}^\infty \sum_{m=0}^\infty \sum_{k=m}^{m+1} g_1 g_2 \sqrt{\eta_1 \eta_2} \chi_1^{2n+1} \alpha_1^{2n} \chi_2^{2m+1} \alpha_2^{2m} (m+1)\sqrt{n+1}\frac{(-\gamma)^{m+k}}{k!}\gamma^{k-m} \ket{n,0}\bra{n+1,1}_{AB}
    \\
    &\quad + \sum_{n=0}^\infty \sum_{m=1}^\infty g2 \sqrt{\eta_2}\chi_1^{2n} \alpha_1^{2n} \chi_2^{2m - 1} \alpha_2^{2m - 2} \frac{(-\gamma)^{2m-1}}{(m-1)!} \ket{n,1}\bra{n,0}_{AB}
    \\
    &\quad + \sum_{n=0}^\infty \sum_{m=1}^\infty \ g_2^2 \eta_2 \chi_1^{2n} \alpha_1^{2n} \chi_2^{2m} \alpha_2^{2m - 2} \frac{(-\gamma)^{2m}}{(m-1)!} \ket{n,1}\bra{n,1}_{AB}
    \\
    &\quad + \sum_{n=0}^\infty \sum_{m=1}^\infty \sum_{k=\mathrm{max}(m-2,0)}^{m-1} g_1 g_2 \sqrt{\eta_1 \eta_2} \chi_1^{2n+1} \alpha_1^{2n} \chi_2^{2m-1} \alpha_2^{2m - 2} \sqrt{n+1} \frac{(-\gamma)^{m+k}}{k!}\gamma^{2+k-m} \ket{n,1}\bra{n+1,0}_{AB}
    \\
    &\quad + \sum_{n=0}^\infty \sum_{m=1}^\infty \sum_{k=\mathrm{max}(m-1,0)}^m g_1 g_2^2 \sqrt{\eta_1} \eta_2 \chi_1^{2n+1} \alpha_1^{2n} \chi_2^{2m} \alpha_2^{2m - 2} m\sqrt{n+1}\frac{(-\gamma)^{m+k}}{k!}\gamma^{1+k-m} \ket{n,1}\bra{n+1,1}_{AB}
    \\
    &\quad + \sum_{n=1}^\infty \sum_{m=0}^\infty \sum_{k=\mathrm{max}(m-1,0)}^m g_1 \sqrt{\eta_1}\chi_1^{2n-1} \alpha_1^{2n - 2} \chi_2^{2m} \alpha_2^{2m} \sqrt{n}\frac{(-\gamma)^{m+k}}{k!} \gamma^{1+k-m} \ket{n,0}\bra{n-1,0}_{AB}
    \\
    &\quad + \sum_{n=1}^\infty \sum_{m=0}^\infty \sum_{k=\mathrm{max}(m-1,0)}^m g_1 g_2 \sqrt{\eta_1 \eta_2}\chi_1^{2n-1} \alpha_1^{2n - 2} \chi_2^{2m+1} \alpha_2^{2m} \sqrt{n}\frac{(-\gamma)^{m+k+1}}{k!} \gamma^{1+k-m} \ket{n,0}\bra{n-1,1}_{AB}
    \\
    &\quad + \sum_{n=1}^\infty \sum_{m=0}^\infty \sum_{k=\mathrm{max}(m-1,0)}^m \sum_{j=\mathrm{max}(m-1,0)}^m g_1^2 \eta_1\chi_1^{2n} \alpha_1^{2n - 2} \chi_2^{2m} \alpha_2^{2m} n m! \gamma^{2+k+j-2m} \frac{(-\gamma)^{k+j}}{k!j!}\ket{n,0}\bra{n,0}_{AB}
    \\
    &\quad + \sum_{n=1}^\infty \sum_{m=0}^\infty \sum_{k=\mathrm{max}(m-1,0)}^m \sum_{j=m}^{m+1} g_1^2 g_2 \eta_1 \sqrt{\eta_2} \chi_1^{2n} \alpha_1^{2n - 2} \chi_2^{2m+1} \alpha_2^{2m} n (m+1)! \gamma^{1+k+j-2m} \frac{(-\gamma)^{k+j}}{k!j!} \ket{n,0}\bra{n,1}_{AB}
    \\
    &\quad + \sum_{n=1}^\infty \sum_{m=1}^\infty \sum_{k=m-1}^m g_1 g_2 \sqrt{\eta_1 \eta_2}\chi_1^{2n-1} \alpha_1^{2n - 2} \chi_2^{2m-1} \alpha_2^{2m - 2} \sqrt{nm}\frac{(-\gamma)^{m+k-1}}{k!} \gamma^{1+k-m} \ket{n,1}\bra{n-1,0}_{AB}
    \\
    &\quad + \sum_{n=1}^\infty \sum_{m=1}^\infty \sum_{k=m-1}^m g_1 g_2^2 \sqrt{\eta_1} \eta_2\chi_1^{2n-1} \alpha_1^{2n - 2} \chi_2^{2m} \alpha_2^{2m - 2} \sqrt{n} m \frac{(-\gamma)^{m+k}}{k!} \gamma^{1+k-m} \ket{n,1}\bra{n-1,1}_{AB}
    \\
    &\quad + \sum_{n=1}^\infty \sum_{m=1}^\infty \sum_{k=m-1}^m \sum_{j=\mathrm{max}(m-2,0)}^{m-1} g_1^2 g_2 \eta_1 \sqrt{\eta_2} \chi_1^{2n} \alpha_1^{2n - 2} \chi_2^{2m-1} \alpha_2^{2m - 2} n m! \gamma^{3+k+j-2m} \frac{(-\gamma)^{k+j}}{k!j!} \ket{n,1}\bra{n,0}_{AB}
    \\
    &\quad + \sum_{n=1}^\infty \sum_{m=1}^\infty \sum_{k=m-1}^m \sum_{j=m-1}^{m} g_1^2 g_2^2 \eta_1 \eta_2 \chi_1^{2n} \alpha_1^{2n - 2} \chi_2^{2m} \alpha_2^{2m - 2} n m m! \gamma^{2+k+j-2m} \frac{(-\gamma)^{j+k}}{k!j!} \ket{n,1}\bra{n,1}_{AB} \Bigg].
\end{aligned}
\end{eqnarray}
\end{widetext}

\noindent By then further tracing out Bob's mode, $B$, we find Alice's conditional density operator $\rho_A(\gamma)$. Using $\rho_{AB}(\gamma)$ and $\rho_{A}(\gamma)$, we calculate the reverse coherent information (RCI) between Alice and Bob as
\begin{align}
    I_{AB}(\gamma) = H(\rho_A) - H(\rho_{AB}),\ \ H(\rho) \equiv -\sum \lambda_i \mathrm{log}_2 \lambda_i
\end{align}

\noindent where $\lambda_i$ are the eigenvalues of $\rho$. 
In the integration window $|\gamma| < \gamma_{\mathrm{max}}$, the conditional entanglement distribution rate $R(\gamma)$ is then calculated by multiplying the RCI of the conditional end-to-end state $I_{AB}(\gamma)$ by the multiplexing-boosted success probability of quantum scissor-based probabilistic noiseless linear amplification succeeding on at least one half channel on either side of the repeater node.
In other words, for $|\gamma| < \gamma_{\mathrm{max}}$, 
\begin{align}
    R(\gamma) &= \frac{(1 - (1 - P_{NLA,1})^M)(1 - (1 - P_{NLA,2})^M)}{M}\nonumber\\
    &\times \mathrm{max}(I_{AB}(\gamma),0).
\end{align}

\noindent We get the asymptotic ergodic entanglement distribution rate by averaging $R(\gamma)$ over our window $|\gamma| < \gamma_{\mathrm{max}}$:

\begin{align}
\bar{R} &= \int_{|\gamma| < \gamma_{\mathrm{max}}} R(\gamma) P_{\Gamma}(\gamma) \mathrm{d}^2\gamma\nonumber\\
&= \frac{2\pi \int_0^{\gamma_{\mathrm{max}}} R(\gamma) \mathrm{Tr}_{AB}[\rho_{AB}(\gamma)] \gamma \mathrm{d}\gamma}{2\pi \int_0^\infty \mathrm{Tr}_{AB}[\rho_{AB}(\gamma)] \gamma \mathrm{d}\gamma}
\end{align}

\noindent noting that, just like in Ref.~\cite{diasCVRepeater}, we set up $\ket{\gamma}_{FC}$ so that the trace of $\rho_{AB}$ corresponds to $P_{\Gamma}(\gamma)$, an unnormalized probability distribution of $\gamma$. $\gamma_{\mathrm{max}}$ is chosen such that 

$$\int_{\gamma < |\gamma_{\mathrm{max}}|} P_\Gamma(\gamma) d^2\gamma = \int_0^{\gamma_\mathrm{max}} P_\Gamma(\gamma) 2 \pi \gamma d\gamma \geq 0.99.$$ 

\noindent We choose to not integrate over the entire complex plane for computational simplicity. Both the RCI and $P_\Gamma (\gamma)$ vanish with larger $|\gamma|$, with RCI eventually going negative, so in every scenario the RCI lost to not averaging over the entire complex plane is negligible if not zero.

Similarly, for the symmetric case where TMSV sources are at Alice and Bob's end, we can do almost identical algebra to arrive at the following (unnormalized) joint density matrix (also expanded in the Fock basis):

\begin{widetext}

\begin{eqnarray}
\begin{aligned}
    \rho_{AB} &= \frac{e^{-\gamma^2}}{\pi} \left( \frac{1-\chi_1^2}{1+g_1^2} \right) \left( \frac{1-\chi_2^2}{1+g_2^2} \right)
    \\
    &\quad \Bigg[ \sum_{n=0}^\infty \sum_{m=0}^\infty \chi_1^{2n} \alpha_1^{2n} \chi_2^{2m} \alpha_2^{2m} \ket{n,m}\bra{n,m}_{AB}
    \\
    &\quad + \sum_{n=0}^\infty \sum_{m=0}^\infty g_2 \sqrt{\eta_2} \chi_1^{2n} \alpha_1^{2n} \chi_2^{2m+1} \alpha_2^{2m} (-\gamma) \sqrt{m+1} \ket{n,m}\bra{n,m+1}_{AB}
    \\
    &\quad + \sum_{n=0}^\infty \sum_{m=0}^\infty  g_1 \sqrt{\eta_1} \chi_1^{2n+1} \alpha_1^{2n} \chi_2^{2m} \alpha_2^{2m} \sqrt{n+1} \gamma \ket{n,m}\bra{n+1,m}_{AB}
    \\
    &\quad + \sum_{n=0}^\infty \sum_{m=0}^\infty  g_1 g_2 \sqrt{\eta_1 \eta_2} \chi_1^{2n+1} \alpha_1^{2n} \chi_2^{2m+1} \alpha_2^{2m} \sqrt{(m+1)(n+1)}(1-|\gamma|^2) \ket{n,m}\bra{n+1,m+1}_{AB}
    \\
    &\quad + \sum_{n=0}^\infty \sum_{m=1}^\infty g2 \sqrt{\eta_2}\chi_1^{2n} \alpha_1^{2n} \chi_2^{2m - 1} \alpha_2^{2m - 2} \sqrt{m} (-\gamma) \ket{n,m}\bra{n,m-1}_{AB}
    \\
    &\quad + \sum_{n=0}^\infty \sum_{m=1}^\infty \ g_2^2 \eta_2 \chi_1^{2n} \alpha_1^{2n} \chi_2^{2m} \alpha_2^{2m - 2} m |\gamma|^2  \ket{n,m}\bra{n,m}_{AB}
    \\
    &\quad + \sum_{n=0}^\infty \sum_{m=1}^\infty g_1 g_2 \sqrt{\eta_1 \eta_2} \chi_1^{2n+1} \alpha_1^{2n} \chi_2^{2m-1} \alpha_2^{2m - 2} \sqrt{m(n+1)} (-\gamma^2) \ket{n,m}\bra{n+1,m-1}_{AB}
    \\
    &\quad + \sum_{n=0}^\infty \sum_{m=1}^\infty  g_1 g_2^2 \sqrt{\eta_1} \eta_2 \chi_1^{2n+1} \alpha_1^{2n} \chi_2^{2m} \alpha_2^{2m - 2} m\sqrt{n+1} (-\gamma)(1-|\gamma|^2) \ket{n,m}\bra{n+1,m}_{AB}\nonumber
    \end{aligned}
\end{eqnarray}
\begin{eqnarray}
\begin{aligned}
    &\quad + \sum_{n=1}^\infty \sum_{m=0}^\infty  g_1 \sqrt{\eta_1}\chi_1^{2n-1} \alpha_1^{2n - 2} \chi_2^{2m} \alpha_2^{2m} \sqrt{n}\gamma^* \ket{n,m}\bra{n-1,m}_{AB}
    \\
    &\quad + \sum_{n=1}^\infty \sum_{m=0}^\infty g_1 g_2 \sqrt{\eta_1 \eta_2}\chi_1^{2n-1} \alpha_1^{2n - 2} \chi_2^{2m+1} \alpha_2^{2m} \sqrt{n(m+1)}(-\gamma^{*2}) \ket{n,m}\bra{n-1,m+1}_{AB}
    \\
    &\quad + \sum_{n=1}^\infty \sum_{m=0}^\infty g_1^2 \eta_1\chi_1^{2n} \alpha_1^{2n - 2} \chi_2^{2m} \alpha_2^{2m} n |\gamma|^2\ket{n,m}\bra{n,m}_{AB}
    \\
    &\quad + \sum_{n=1}^\infty \sum_{m=0}^\infty   g_1^2 g_2 \eta_1 \sqrt{\eta_2} \chi_1^{2n} \alpha_1^{2n - 2} \chi_2^{2m+1} \alpha_2^{2m} n \sqrt{(m+1)} \gamma^* (1-|\gamma|^2) \ket{n,m}\bra{n,m+1}_{AB}
    \\
    &\quad + \sum_{n=1}^\infty \sum_{m=1}^\infty  g_1 g_2 \sqrt{\eta_1 \eta_2}\chi_1^{2n-1} \alpha_1^{2n - 2} \chi_2^{2m-1} \alpha_2^{2m - 2} \sqrt{nm} (1-|\gamma|^2) \ket{n,m}\bra{n-1,m-1}_{AB}
    \\
    &\quad + \sum_{n=1}^\infty \sum_{m=1}^\infty  g_1 g_2^2 \sqrt{\eta_1} \eta_2\chi_1^{2n-1} \alpha_1^{2n - 2} \chi_2^{2m} \alpha_2^{2m - 2} \sqrt{n} m (-\gamma^*) (1-|\gamma|^2) \ket{n,m}\bra{n-1,m}_{AB}
    \\
    &\quad + \sum_{n=1}^\infty \sum_{m=1}^\infty  g_1^2 g_2 \eta_1 \sqrt{\eta_2} \chi_1^{2n} \alpha_1^{2n - 2} \chi_2^{2m-1} \alpha_2^{2m - 2} n \sqrt{m} (1-|\gamma|^2)\gamma \ket{n,m}\bra{n,m-1}_{AB}
    \\
    &\quad + \sum_{n=1}^\infty \sum_{m=1}^\infty  g_1^2 g_2^2 \eta_1 \eta_2 \chi_1^{2n} \alpha_1^{2n - 2} \chi_2^{2m} \alpha_2^{2m - 2} n m (1-|\gamma|^2)^2 \ket{n,m}\bra{n,m}_{AB} \Bigg].
\end{aligned}
\end{eqnarray}
\end{widetext}

For the final case, symmetric distribution with the QSs at Alice and Bob's nodes, we use covariance matrix algebra to simplify the scenario. We can use the algebra from \cite{covarianceMatrixAlgebra} to perform a bell state measurement on one mode from each TMSV dual state, leaving a two mode state that can then be propagated through the pure loss channel to reach the NLAs at Alice and Bob. It can be shown that, before the pure loss channel, the resulting two mode state is identical to a TMSV with squeezing parameter $\chi' = \chi_1 \chi_2$. After going through the QSs the resulting unnormalized density matrix is:

\begin{widetext}
\begin{equation}
\begin{aligned}
\rho_{AB} = \begin{pmatrix}
\frac{1}{1 - \alpha_1^2 \alpha_2^2 \chi'^2} & 0 & 0 & \frac{g_1 g_2 \sqrt{\eta_1 \eta_2} \chi'}{(1 - \alpha_1^2 \alpha_2^2 \chi'^2)^2}\\
0 & \frac{g_2^2 \alpha_1^2 \eta_2 \chi'^2}{(1 - \alpha_1^2 \alpha_2^2 \chi'^2)^2} & 0 & 0\\
0 & 0 & \frac{g_1^2 \alpha_2^2 \eta_1 \chi'^2}{(1 - \alpha_1^2 \alpha_2^2 \chi'^2)^2} & 0\\
\frac{g_1 g_2 \sqrt{\eta_1 \eta_2} \chi'}{(1 - \alpha_1^2 \alpha_2^2 \chi'^2)^2} & 0 & 0 & \frac{g_1^2 g_2^2 \alpha_1^2 \eta_1 \eta_2 \chi'^2 (1 + \alpha_1^2 \alpha_2^2 \chi'^2)}{(1 - \alpha_1^2 \alpha_2^2 \chi'^2)^3}
\end{pmatrix}
\end{aligned}
\end{equation}
\end{widetext}

\end{document}